\documentclass[12pt]{article}
\usepackage{graphicx}

\def\h{\eta}

\def\j{\psi}

\def\l{\lambda}
\def\m{\mu}
\def\n{\nu}

\def\p{\pi}

\def\s{\sigma}

\def\O{\Omega}

\def\Q{\Theta}
\def\S{\Sigma}

\def\beq{\begin{equation}}
\def\eeq{\end{equation}}
\def\bea{\begin{eqnarray}}
\def\eea{\end{eqnarray}}

\def\NO{\nonumber}

\def\pl#1#2#3{Phys.~Lett.~{\bf B {#1}} ({#2}) #3}
\def\np#1#2#3{Nucl.~Phys.~{\bf B {#1}} ({#2}) #3}
\def\prl#1#2#3{Phys.~Rev.~Lett.~{\bf #1} ({#2}) #3}
\def\pr#1#2#3{Phys.~Rev.~{\bf D {#1}} ({#2}) #3}

\def\ap#1#2#3{Ann.~of Phys.~{\bf {#1}} ({#2}) #3}

\def\ptp#1#2#3{Progr.~Theor.~Phys.~{\bf {#1}} ({#2}) #3}

\begin{document}
\date{\mbox{ }}
\title{{\normalsize DESY 02-046\hfill\mbox{}\\
April 2002\hfill\mbox{}}\\
\vspace{2cm} \textbf{Exceptional Coset Spaces and\\ 
Unification in Six Dimensions}\\
[8mm]}
\author{T.~Asaka, W.~Buchm\"uller, L.~Covi\\
\textit{Deutsches Elektronen-Synchrotron DESY, Hamburg, Germany}}
\maketitle

\thispagestyle{empty}

\begin{abstract}
\noindent
The coset spaces E$_8$/SO(10)$\times$H$_F$
allow complex structures which can account for three quark-lepton 
generations including right-handed neutrinos. We show that in the context 
of supersymmetric SO(10) gauge theories in 6 dimensions they also provide 
the Higgs fields which are needed to break the electroweak and 
${B-L}$ gauge symmetries, and to generate   
small neutrino masses via the seesaw mechanism.
\end{abstract}

\newpage
The standard model gauge groups of electroweak and strong interactions are 
naturally
unified in the simple gauge group SU(5) \cite{gg74}. With the increasing
experimental evidence for neutrino masses and mixings the larger gauge
group SO(10) \cite{gfm75} with the Pati-Salam subgroup 
SU(4)$\times$SU(2)$\times$SU(2) \cite{ps74}
appears particularly attractive, since all quarks and leptons
of a single generation, including the right-handed neutrino, are then
also unified in a single multiplet.

The unification groups SU(5)$=$E$_4$, SO(10)$=$E$_5$ and E$_6$ \cite{grs75}
belong to the sequence of exceptional groups E$_n$ which terminates at 
E$_8$. This largest exceptional group is
attractive for unification \cite{baa80,bg80} since its smallest multiplet, 
the 248-dimensional adjoint representation, is large enough to accommodate
all three generations of quarks and leptons. The theory is naturally
supersymmetric. However, in addition to the three known quark-lepton
generations the theory also predicts three light mirror generations, 
contrary to observation.

The group E$_8$ also appears in the ten-dimensional Yang-Mills supergravity
theory where a cancellation of all gauge and gravitational anomalies is
obtained by means of the Green-Schwarz mechanism for the group E$_8\times$E$_8$
\cite{gs84}. Compactification to four dimensions \cite{chx85,dhx85}
can yield low energy effective theories with unbroken $N=1$ supersymmetry 
and chiral fermions, similar to the structure of the standard model. The
number of families is determined by the Euler characteristic of the compact 
manifold.

Further, the group  E$_8$ has been considered in attempts to relate quarks 
and leptons to a coset space G/H, where G is an appropriate simple group and 
H contains the standard model gauge group \cite{blx82}. By pairing the scalar
degrees of freedom of G/H into complex fields, which become the superpartner
of quarks and leptons, the problem of mirror families can be
avoided \cite{ong83}.
Particularly attractive are coset spaces E$_8$/(SO(10)$\times$H$_F$) where 
H$_F$ is a subgroup of SU(3)$\times$U(1) \cite{ong83}-\cite{ikk86}. In the
case H$_F =$SU(3)$\times$U(1) the representation of chiral multiplets is
unique,
\beq\label{complex}
\O = ({\bf 16},{\bf 3})_1 + ({\bf 16^*},{\bf 1})_3 
+ ({\bf 10},{\bf 3^*})_2 + ({\bf 1},{\bf 3})_4\;.
\eeq
Hence, in addition to three quark-lepton generations contained in the three
{\bf 16}'s of SO(10), one mirror generation, {\bf 16$^*$}, occurs, again in
contrast to observation. For H$_F$ = SU(2)$\times$U(1)$^2$ and H$_F$ = U(1)$^3$
various complex structures are possible \cite{bn85}, which contain
either three {\bf 16}'s and one {\bf 16$^*$}, or two {\bf 16}'s and two 
{\bf 16$^*$}'s in addition to three {\bf 10}'s and SO(10) singlets.
 
Orbifold compactifications \cite{dhx85} have recently been applied to
GUT field theories. The breaking of the GUT symmetry then automatically
yields the required doublet-triplet splitting of Higgs fields \cite{kaw00}.
Several SU(5) models have been constructed in 5 dimensions 
\cite{kaw00}-\cite{hmr01}, whereas 6 dimensions are required for the 
breaking of SO(10) \cite{abc01,hnx02}. The N=2 supersymmetric SO(10)
theory in 6 dimensions can
also be used as a starting point to obtain a SU(5) GUT with three generations
\cite{wy02}.

In the following we want to point out that the complex structure 
(\ref{complex}), when combined with the orbifold compactification of the 6d
SO(10) theory, yields in a simple way the supersymmetric standard model with
right-handed neutrinos. 
The three {\bf 16}'s, which contain quarks and leptons, are brane fields; the
{\bf 16$^*$} and the three {\bf 10}'s are bulk fields. Vacuum expectation
values of bulk fields break the symmetries
SU(2)$\times$U(1)$_Y$ and U(1)$_X$, which are left unbroken by the orbifold
compactification.

\begin{figure}
\centering 
\includegraphics[scale=0.7]{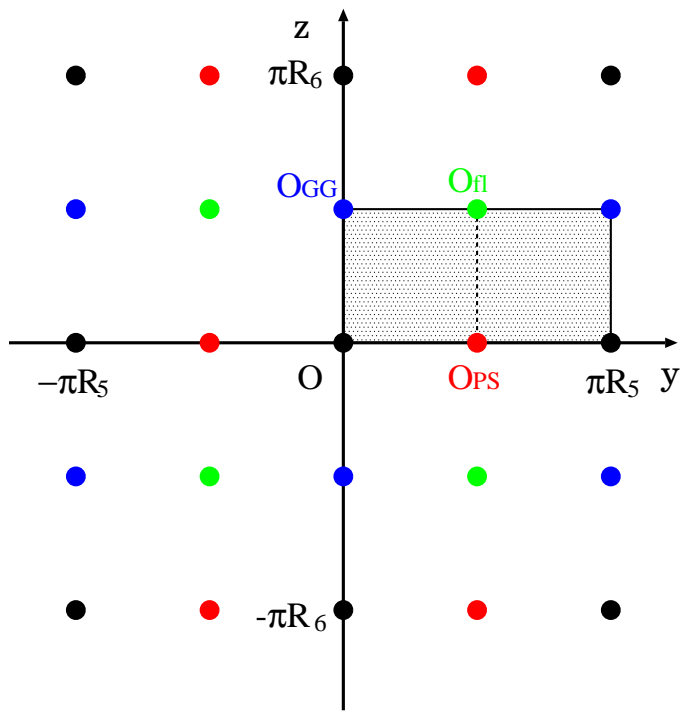}
\caption{Orbifold $T^2/(Z_2^I\times Z_2^{PS}\times Z_2^{GG})$ with the
fixpoints $O$, $O_{PS}$, $O_{GG}$, and $O_{fl}$.\label{fig:orb}}
\end{figure}

Consider the SO(10) gauge theory in 6d with N=1 supersymmetry. The gauge
fields $V_M(x,y,z)$, with $M=\m,5,6$, $x^5=y$, $x^6=z$, and the 
gauginos $\l_1$, $\l_2$ are conveniently grouped into vector and chiral
multiplets of the unbroken N=1 supersymmetry in 4d,
\beq
V = (V_\m,\l_1)\;, \quad \S = (V_{5,6},\l_2)\;.
\eeq 
Here $V$ and $\S$ are matrices in the adjoint representation of SO(10).
Symmetry breaking is achieved by compactification on the
orbifold $T^2/(Z_2^I\times Z_2^{PS}\times Z_2^{GG})$. The discrete symmetries
$Z_2$ break the extended supersymmetry in 4d; 
they also break the SO(10) gauge group down  
to the subgroups SO(10), G$_{PS}$=SU(4)$\times$SU(2)$\times$SU(2) and 
G$_{GG}$=SU(5)$\times$U(1)$_X$, respectively, at three different fixpoints,
\newpage
\bea\label{fixp}
P_IV(x,-y,-z)P_I^{-1} &=& \h_I V(x,y,z)\;,\\
P_{PS}V(x,-y+{\p R_5/2},-z)P_{PS}^{-1} &=& 
\h_{PS} V(x,y+{\p R_5/2},z)\;,\\
P_{GG}V(x,-y,-z+{\p R_6/2})P_{GG}^{-1} &=& 
\h_{GG} V(x,y,z+{\p R_6/2})\;.
\eea 
Here $P_I=I$, the matrices $P_{PS}$ and $P_{GG}$ are given in 
ref.~\cite{abc01}, and the parities are chosen as $\h_I=\h_{PS}=\h_{GG}=+1$. 
The extended supersymmetry
is broken by choosing in the corresponding equations for $\S$  all parities 
$\h_i=-1$. 
At the fixpoints in the extra dimensions,
$O=(0,0)$, $O_{PS}=(\p R_5/2,0)$ and $O_{GG}=(0,\p R_6/2)$ the
unbroken subgroups are SO(10), G$_{PS}$ and G$_{GG}$, respectively. In 
addition, there is a fourth fixpoint at  $O_{fl}=(\p R_5/2,\p R_6/2)$ 
\cite{hnx02}, which is obtained by combining the three discrete symmetries
$Z_2$, $Z_2^{PS}$ and $Z_2^{GG}$ defined above,
\beq
P_{fl}V(x,-y +{\p R_5/2},-z+{\p R_6/2})P_{fl}^{-1} = 
+ V(x,y+{\p R_5/2},z+{\p R_6/2})\;.
\eeq 
The unbroken subgroup at the fixpoint $O_{fl}$ is flipped SU(5), i.e.
G$_{fl}$=SU(5)$'\times$U(1)$'$. The physical region is obtained by
folding the shaded regions in fig.~\ref{fig:orb} along the dotted line and 
gluing the edges. The result is a `pillow' with the four fixpoints as corners. 
The unbroken gauge group of the effective 4d theory
is given by the intersection of the SO(10) subgroups at the fixpoints. In
this way one obtains the standard model group with an additional U(1) factor,
G$_{SM'}$= SU(3)$\times$SU(2)$\times$U(1)$_Y \times$U(1)$_X$. The difference of
baryon and lepton number is the linear combination
$B-L = \sqrt{{16\over 15}}Y-\sqrt{{8\over 5}}X$. The zero modes of 
the vector multiplet $V$ form the gauge fields of G$_{SM'}$.

The vector multiplet $V$ is a {\bf 45}-plet of SO(10) which has an
irreducible anomaly in 6 dimensions. It is related to the irreducible
anomalies of hypermultiplets in the fundamental and the spinor 
representations by (cf.~\cite{hmr02}),
\beq
a({\bf 45}) = - 2 a({\bf 10})\;, \quad 
a({\bf 16}) = a({\bf 16^*}) = - a({\bf 10})\;.
\eeq
Hence, the anomaly of the vector multiplet can be cancelled by adding two
{\bf 10} hypermultiplets, $H_1$ and $H_2$. The cancellation of the reducible
anomalies can be achieved by means of the Green-Schwarz mechanism \cite{gs84}.

For these hypermultiplets we have to define parities with respect to the
discrete symmetries,
\bea
P_I H(x,-y,-z) &=& \h_I H(x,y,z)\;,  \\
P_{PS}H(x,-y+\p R_5/2,-z) &=& \h_{PS} H(x,y+\p R_5/2,z)\;,\\
P_{GG}H(x,-y,-z+\p R_6/2) &=& \h_{GG} H(x,y,z+\p R_6/2)\;,  
\eea
with $\h_i= \pm 1$ ($i=I,PS,GG$). All hypermultiplets split under the extended 
6d supersymmetry into two N=1 4d chiral multiplets, $H = (H,H')$. They have 
the 6d superpotential interactions \cite{agw01},
\beq
S_6 = \int d^6x \left(\int d^2 \Q H'(\partial + \sqrt{2} g \S)H 
+ h.c. \right)\;,
\eeq
where $\partial = \partial_5 - i\partial_6$ and $g$ is the 6d gauge coupling. 
Invariance of the action then requires that the parities of $H$ and $H'$
are opposite. In the following we denote by $\h_i$ the parities of the first
4d chiral multiplet, and we choose $\h_I = +1$.

\renewcommand{\arraystretch}{1.2}
\begin{table}
  \begin{center}
   $\begin{array}[h]{|c||cc|cc|cc|cc|}\hline
     \mbox{SO(10)} &
     \multicolumn{8}{|c|}{ \mathbf{10} }
     \\ \hline
     G_{PS} &
     \multicolumn{2}{|c|}{ ( \mathbf{1}, \mathbf{2}, \mathbf{2}) } &
     \multicolumn{2}{|c|}{ ( \mathbf{1}, \mathbf{2}, \mathbf{2}) } &
     \multicolumn{2}{|c|}{ ( \mathbf{6}, \mathbf{1}, \mathbf{1}) } &
     \multicolumn{2}{|c|}{ ( \mathbf{6}, \mathbf{1}, \mathbf{1}) }
     \\ \hline
     G_{GG} &
     \multicolumn{2}{|c|}{ \mathbf{5}^\ast{}_{-2} } &
     \multicolumn{2}{|c|}{ \mathbf{5}{}_{+2} } &
     \multicolumn{2}{|c|}{ \mathbf{5}^\ast{}_{-2} }  &
     \multicolumn{2}{|c|}{ \mathbf{5}{}_{+2} }
     \\ \hline
        &  \multicolumn{2}{|c|}{H^c} & \multicolumn{2}{|c|}{H} &
     \multicolumn{2}{|c|}{G^c} & \multicolumn{2}{|c|}{G}
     \\
     {} &
    Z_2^{PS} & Z_2^{GG} &
    Z_2^{PS} & Z_2^{GG} &
    Z_2^{PS} & Z_2^{GG} &
    Z_2^{PS} & Z_2^{GG}
    \\ \hline \hline
     H_1 &
     + & + &
     + & - &
     - & + &
     - & -
     \\ \hline
     H_2 &
     + & - &
     + & + &
     - & - &
     - & +
     \\ \hline
     H_3 &
     - & + &
     - & - &
     + & + &
     + & -
     \\ \hline
    \end{array}$
    \caption{Parity assignment for the bulk $\mathbf{10}$ hypermultiplets.
      $H_1^c = H_d$ and $H_2 = H_u$.}
    \label{tab:P10}
  \end{center}
\end{table}

The discrete symmetry $Z_{PS}$ implies automatically a splitting between
the SU(2) doublets and the SU(3) triplets contained in the {\bf 10}-plets.
The choice $\h_{PS} = +1$ leads to massless SU(2) doublets and massive
colour triplets (cf.~table~\ref{tab:P10}). Choosing further $\h_{GG} = +1$ for
$H_1$ and $\h_{GG} = -1$ for $H_2$, selects the doublet $H^c$ from the SU(5)
{\bf 5$^*$}-plet contained in $H_1$, and the doublet $H$ from the SU(5) 
{\bf 5}-plet of $H_2$ (cf.~table~\ref{tab:P10}). The doublets $H^c$ and $H$ have
the quantum numbers of the Higgs fields $H_d$ and $H_u$ in the supersymmetric
standard model.

Quarks and leptons can be incorporated by adding {\bf 16}-plets,
additional {\bf 10}-plets etc. in the bulk and on the fixpoints. Without any
constraint on the multiplicity of these fields there are many possibilities
\cite{hnx02,hks01}. It is remarkable that for the complex structure $\O$ given
in eq.~(1), the requirement of SO(10) anomaly cancellations in the bulk 
determines the distribution of multiplets uniquely. As already mentioned,
two {\bf 10}'s are required to cancel the anomaly of the {\bf 45} vector
multiplet. Hence, the ({\bf 10},{\bf 3$^*$}) have to be bulk fields. The anomaly 
of the third {\bf 10}-plet can only be cancelled by the ({\bf 16$^*$},{\bf 1}),
which leaves ({\bf 16},{\bf 3}) as brane fields. Note, that in general a bulk field 
contains two 4-dimensional $N=1$ chiral multiplets which transform as complex 
conjugates of each other with respect to the
gauge group. Hence, in our SO(10) model in 6 dimensions the content of 4d chiral
multiplets is larger than the complex structure (\ref{complex}).

\renewcommand{\arraystretch}{1.2}
\begin{table}
  \begin{center}
  $\begin{array}[h]{|c||cc|cc|cc|cc|}\hline
    \mbox{SO(10)} & \multicolumn{8}{|c|}{ \mathbf{16}^\ast }
    \\ \hline
    G_{PS} &
    \multicolumn{2}{|c|}{ (\mathbf{4}^\ast, \mathbf{2}, \mathbf{1}) } &
    \multicolumn{2}{|c|}{ (\mathbf{4}^\ast, \mathbf{2}, \mathbf{1}) } &
    \multicolumn{2}{|c|}{ (\mathbf{4}, \mathbf{1}, \mathbf{2}) }  &
    \multicolumn{2}{|c|}{ (\mathbf{4}, \mathbf{1}, \mathbf{2}) }
    \\ \hline
    G_{GG} &
    \multicolumn{2}{|c|}{ \mathbf{10}^\ast{}_{+1} } &
    \multicolumn{2}{|c|}{ \mathbf{5}_{-3} } &
    \multicolumn{2}{|c|}{ \mathbf{10}^\ast{}_{ +1} } &
    \multicolumn{2}{|c|}{ \mathbf{5}_{-3},  \mathbf{1}{}_{+5} }
    \\ \hline
    {} &
    \multicolumn{2}{|c|}{ Q^c} &
    \multicolumn{2}{|c|}{L^c} &
    \multicolumn{2}{|c|}{U^c, E^c} &
    \multicolumn{2}{|c|}{D, N}
    \\
    {} &
    Z_2^{PS} & Z_2^{GG} &
    Z_2^{PS} & Z_2^{GG} &
    Z_2^{PS} & Z_2^{GG} &
    Z_2^{PS} & Z_2^{GG}
     \\ \hline
     \Phi^c &
     - & - &
     - & + &
     + & - &
     + & +
     \\ \hline
    \end{array}$
    \caption{Parity assignment for the bulk $\mathbf{16}^\ast$ hypermultiplet.}
    \label{tab:P16b}
  \end{center}
\end{table}

We now have to choose the parities for $H_3$, the third {\bf 10}-plet, and for
$\Phi^c$, the {\bf 16$^*$}-plet. The presence of $\Phi^c$ offers the possibility
to break U(1)$_{B-L}$ spontaneously by a vacuum expectation value of its
SU(5) singlet component $N$ (table~\ref{tab:P16b}). To have $N$ as zero mode
fixes the parities of $\Phi^c$ to be $\h_{PS}=\h_{GG}=-1$. Then an additional
colour triplet, $D$, appears as zero mode. As we shall see, $D$ can acquire
a Dirac mass together with another colour triplet, $G^c$, which can be
chosen as zero mode of the third {\bf 10}-plet. The corresponding parities
of $H_3$ are $\h_{PS}=-1$, $\h_{GG}=+1$. The parities of all components
of $H_i$ and $\Phi^c$ are listed in tables~\ref{tab:P10} and \ref{tab:P16b}.

Knowing the parities of all fields we can now discuss the superpotential.
We consider the three brane fields {\bf 16}, $\j_i$, on the SO(10) symmetric
fixpoint $O$. To restrict the number of terms we require R-invariance 
and an additional global U(1)$_{\tilde{X}}$ symmetry (cf.~table~\ref{tab:rpa}). 
The most general superpotential up to quartic terms is then given by, 
\bea\label{O10}
W_4 &=&  h_d \j \j H_1 + h_u \j \j H_2 + {h_N\over M_*} \j \j \Phi^c \Phi^c \NO\\
&& + M_1 H_1 H_3 + M_2 H_2 H_3 + \l_1 \Phi^c \Phi^c H_3 \;,
\eea
where we choose $M_* > 1/R_{5,6}$ to be the cutoff of the 6d theory, and the
bulk fields have been properly normalised. The first three terms
are familiar from ordinary SO(10) GUTs whereas the last three terms are
additional couplings among bulk fields. 

\renewcommand{\arraystretch}{1.2}
\begin{table}[b]
  \begin{center}
    $\begin{array}[h]{| c | c c c c c |}
      \hline
      {} & H_1 & H_2 & \psi_i & \Phi^c & H_3
      \\ \hline
      R &  0 & 0 & 1 & 0 & 2
   \\ \hline
      \widetilde{X} & - 2 a & - 2 a & a & - a & 2 a
   \\ \hline
    \end{array}$
    \caption{Charge assignments for the symmetries U(1)$_R$ and U(1)$_{\tilde{X}}$.}
    \label{tab:rpa}
  \end{center}
\end{table}

It is instructive to consider in eq.~(\ref{O10}) just the zero mode interactions
and their couplings to a single heavy field in the bilinear and cubic terms.
In standard notation, with $\j = (q,u^c, e^c,l,d^c,n^c)$, one obtains,
\bea
W_4 &=& h_d (q d^c + l e^c) H_1^c + h_u (q u^c + l n^c) H_2 \NO\\
&& +{h_N\over M_*} (n^c N)^2 
+  {h_N'\over M_*} (d^c D)^2
+  {h_N''\over M_*} d^c D n^c N  \NO\\
&& + \l_1 D N G_3^c + M_1 H_1^c H_3 + M_2 H_2 H_3^c \NO\\
&& +h_d (q u^c + l n^c) H_1 + h_d (q q + u^c e^c + d^c n^c) G_1
   + h_d (q l + u^c d^c) G_1^c \NO\\
&& +h_u (q d^c + l e^c) H_2^c + h_u (q q + u^c e^c + d^c n^c) G_2 
   + h_u (q l + u^c d^c) G_2^c\;;
\eea
here the three couplings $h_N$, $h_N'$ and $h_N''$ correspond to the three SO(10)
invariants which can be formed from $\j \j \Phi^c \Phi^c$.

Vacuum expectation values $v_1 = \langle H_1^c\rangle$,
$v_2 = \langle H_2 \rangle$ and $v_N = \langle N \rangle$ yield the mass terms,
\beq\label{mass}
W_m = m_u u u^c +  m_\n n n^c + m_d d d^c + m_e e e^c + 
m_N n^c n^c \;,
\eeq
with mass matrices $m_u = m_\n = h_u v_2$, $m_d = m_e = h_d v_1$ and 
$m_N = h_N v_N^2/M_*$. With $v_{1,2} \ll v_N$ this gives, to first approximation,
a good description of the observed properties of quarks and leptons. 
Since there are two Higgs doublets, $H_1$ and $H_2$, the unwanted relation
of minimal SO(10) models, $m_u = m_d$, is avoided. The evidence for small 
neutrino masses, together with the seesaw mechanism \cite{yan79},
suggests that U(1)$_{B-L}$ is broken near the unification scale, 
i.e. $v_N \sim 1/R_{5,6}$, which implies large Majorana masses \cite{wit80}. 
This also leads to a large mass term, $m_T D G^c_3$,
for the colour triplet zero modes of $\Phi^c$ and $H_3$,
with $m_T = \l_1 v_N$. The vacuum expectation value 
$\langle N \rangle =v_N$ breaks U(1)$_{\tilde{X}}\times$U(1)$_X$ to the diagonal 
global subgroup, leaving U(1)$_R$ unbroken. As a consequence, integrating out the
heavy fields $G_{1,2}$, $G_{1,2}^c$ does not lead to a dimension-5 operator
for proton decay. For the same reason, integrating out $H_3$ and $H_3^c$
does not generate a $\m$-term $H_1^c H_2$. Finally, the electroweak scale 
$v_{1,2}$ may be induced together with the $\m$-term
by supersymmetry breaking. The resulting low energy 
effective theory is just the supersymmetric standard model with right-handed
neutrinos. 

\renewcommand{\arraystretch}{1.2}
\begin{table}
\begin{center}
   $\begin{array}[h]{|c||cc|cc|cc|cc|}\hline
     \mbox{SO(10)} &
     \multicolumn{8}{|c|}{ \mathbf{10} }
     \\ \hline
     G_{PS} &
     \multicolumn{2}{|c|}{ ( \mathbf{1}, \mathbf{2}, \mathbf{2}) } &
     \multicolumn{2}{|c|}{ ( \mathbf{1}, \mathbf{2}, \mathbf{2}) } &
     \multicolumn{2}{|c|}{ ( \mathbf{6}, \mathbf{1}, \mathbf{1}) } &
     \multicolumn{2}{|c|}{ ( \mathbf{6}, \mathbf{1}, \mathbf{1}) }
     \\ \hline
     G_{GG} &
     \multicolumn{2}{|c|}{ \mathbf{5}^\ast{}_{-2} } &
     \multicolumn{2}{|c|}{ \mathbf{5}{}_{+2} } &
     \multicolumn{2}{|c|}{ \mathbf{5}^\ast{}_{-2} }  &
     \multicolumn{2}{|c|}{ \mathbf{5}{}_{+2} }
     \\ \hline
        &  \multicolumn{2}{|c|}{H^c} & \multicolumn{2}{|c|}{H} &
     \multicolumn{2}{|c|}{G^c} & \multicolumn{2}{|c|}{G}
     \\
     {} &
    Z_2^{PS} & Z_2^{GG} &
    Z_2^{PS} & Z_2^{GG} &
    Z_2^{PS} & Z_2^{GG} &
    Z_2^{PS} & Z_2^{GG}
    \\ \hline \hline
     H_4 &
     - & - &
     - & + &
     + & - &
     + & +
     \\ \hline
    \end{array}$
  \end{center}
  \begin{center}
  $\begin{array}[h]{|c||cc|cc|cc|cc|}\hline
    \mbox{SO(10)} & \multicolumn{8}{|c|}{ \mathbf{16} }
    \\ \hline
    G_{PS} &
    \multicolumn{2}{|c|}{ (\mathbf{4}, \mathbf{2}, \mathbf{1}) } &
    \multicolumn{2}{|c|}{ (\mathbf{4}, \mathbf{2}, \mathbf{1}) } &
    \multicolumn{2}{|c|}{ (\mathbf{4}^\ast, \mathbf{1}, \mathbf{2}) } &
    \multicolumn{2}{|c|}{ (\mathbf{4}^\ast, \mathbf{1}, \mathbf{2}) }
    \\ \hline
    G_{GG} &
    \multicolumn{2}{|c|}{ \mathbf{10}_{-1} } &
    \multicolumn{2}{|c|}{ \mathbf{5}^\ast{}_{+3} } &
    \multicolumn{2}{|c|}{ \mathbf{10}_{-1 }} &
    \multicolumn{2}{|c|}{ \mathbf{5}^\ast{}_{ +3 }, \mathbf{1}_{ -5 } }
    \\ \hline
    {} &
    \multicolumn{2}{|c|}{ Q}      &
    \multicolumn{2}{|c|}{L}       &
    \multicolumn{2}{|c|}{U, E}  &
    \multicolumn{2}{|c|}{D^c, N^c}
    \\
    {} &
    Z_2^{PS} & Z_2^{GG} &
    Z_2^{PS} & Z_2^{GG} &
    Z_2^{PS} & Z_2^{GG} &
    Z_2^{PS} & Z_2^{GG}
     \\ \hline
     \Phi &
     - & - &
     - & + &
     + & - &
     + & +
     \\ \hline
    \end{array}$
    \caption{Parity assignment for the bulk $\mathbf{10}$ and $\mathbf{16}$
      hypermultiplets.}
    \label{tab:P16}
  \end{center}
\end{table}

What determines the vacuum expectation value $\langle N\rangle$ which breaks
U(1)$_X$ and therefore U(1)$_{B-L}$? Note, that these U(1) symmetries are 
anomalous, since the zero modes of $\Phi^c$ and $H_3$ lead to 4d anomalies 
SU(3)$^2\times$U(1)$_X$ and U(1)$_X^3$.
In general, these can be compensated
by a Chern-Simons term\footnote{For recent discussions in 5-dimensional
theories and references, see \cite{bar02,gno02}}. One may then hope to break 
U(1)$_X$ spontaneously by means of a Fayet-Iliopoulos term on the Georgi-Glashow 
fixpoint $O_{GG}$. This will be discussed in more detail elsewhere \cite{abc02}.

Alternatively, one can avoid the occurrence of anomalies by `partial
doubling', familiar from supersymmetric $\s$-models where one associates
an entire chiral multiplet with some degrees of freedom of the coset space G/H 
(cf.~\cite{bl87}). Let us then add two bulk fields, a {\bf 16}-plet, $\Phi$,
and a fourth {\bf 10}-plet, $H_4$, which have no irreducible 6d anomaly.
The choice of parities $\h_{PS}=-1$ and $\h_{GG}=-1$ for both fields leads
to the zero modes $D^c$, $N^c$ and $G$ (cf.~table~\ref{tab:P16}). The zero modes 
of all bulk fields now
form a real, and therefore anomaly free, representation of $G_{SM'}$.
The additional superpotential terms on the SO(10) brane are,
\bea\label{supdoub}
W_4' &=& M^2 S + \l_2 S \Phi\Phi^c + \l_3 \Phi \Phi H_4\NO\\ 
&& + {\l_4\over M_*} S \Phi\Phi H_1 + {\l_5\over M_*} S \Phi\Phi H_2
   + {\l_6\over M_*} \Phi\Phi^c H_1 H_3 + {\l_7\over M_*} \Phi\Phi^c H_2 H_3\;,
\eea
where we have also included one of the SO(10) singlets, $S$. The $R$- and 
$\tilde{X}$-charges of the additional fields are $R_{\Phi}=0$, $R_{H_4}=R_S=2$,
$\tilde{X}_{\Phi}=a$, $\tilde{X}_{H_4}=-2a$, $\tilde{X}_S=0$. Without the
singlet field $S$, there is a bulk D- and F-flat direction, 
$\langle N \rangle = \langle N^c \rangle = v_N$ \cite{abc02}. The couplings
of $S$ on the brane lift this degeneracy, and one has 
$ v_N = \langle N \rangle = \langle N^c \rangle = \sqrt{-M^2/\l_2}$. 
The cubic term in (\ref{supdoub}) gives mass to two zero modes  $G$ and
$D^c$, $N-N^c$ makes to the U(1)$_X$ vector multiplet massive, and $N+N^c$ has
a common mass term with $S$. Hence,
one obtains the same low energy theory as in the anomalous model discussed above.

We have shown that complex structures of coset spaces 
E$_8$/SO(10)$\times$H$_F$ can provide the starting point of a full grand unified
model in the context of supersymmetric SO(10) theories in 6 dimensions
with orbifold compactification. Open questions concern the breaking of supersymmetry,
the origin of the brane superpotential and, in particular, the possible
connection to theories in 10 dimensions with E$_8$ symmetry.  \\

\noindent
We would like to thank A.~Hebecker, H.~P.~Nilles, M.~Olechowski and C.~Scrucca
for helpful discussions.


\end{document}